**Manuscript Title:**

**Band-edge Bilayer Plasmonic Nanostructure for Surface Enhanced Raman Spectroscopy**

**Complete List of Authors and Affiliations:**

**S. Hamed Shams Mousavi \*, Ali A. Eftekhar \*, Amir H. Atabaki \*†, Ali Adibi \***

\* School of Electrical and Computer Engineering, Georgia Institute of Technology, 778 Atlantic Dr. NW, Atlanta, GA 30332

† Research Laboratory of Electronics, Massachusetts Institute of Technology, 77 Massachusetts Ave, Cambridge, Massachusetts 02139, USA

**Abstract**

Spectroscopic analysis of large biomolecules is critical in a number of applications, including medical diagnostics and label-free biosensing. Recently, it has been shown that Raman spectroscopy of proteins can be used to diagnose some diseases, including a few types of cancer. These experiments have however been performed using traditional Raman spectroscopy and the development of the Surface enhanced Raman spectroscopy (SERS) assays suitable for large biomolecules could lead to a substantial decrease in the amount of specimen necessary for these experiments. We present a new method to achieve high local field enhancement in surface enhanced Raman spectroscopy through the simultaneous adjustment of the lattice plasmons and localized surface plasmon polaritons, in a periodic bilayer nanoantenna array resulting in a high enhancement factor over the sensing area, with relatively high uniformity. The proposed plasmonic nanostructure is comprised of two interacting nanoantenna layers, providing a sharp band-edge lattice plasmon mode and a wide-band localized surface plasmon for the separate enhancement of the pump and emitted Raman signals. We demonstrate the application of the proposed nanostructure for the spectral analysis of large biomolecules by binding a protein (streptavidin) selectively on the hot-spots between the two stacked layers, using a low concentration solution (100 nM) and we successfully acquire its SERS spectrum.

Surface enhanced Raman spectroscopy (SERS), using plasmonic nanoantennas, is a well-established technique for spectral analysis in low concentration limits [1-6]. The recent progress in the spectral assignment and Raman instrumentation has lent itself to a revival of interest in the Raman spectroscopy of proteins and other large biomolecules [7-9]. In applications such as cancer diagnosis [10], it has been shown that invaluable insights could be gained by looking into the Raman spectra of proteins, once deemed to be too complex to extract any meaningful information. Nevertheless, the vast majority of these measurements have been performed using conventional Raman spectroscopy at high concentration levels. Transformation of these experiments to SERS could be pivotal in applications, such as pathogen detection, and real-time spectral analysis in low concentration levels requiring high sensitivity, and it can lead to a substantial decrease in material cost.

A number of plasmonic structures can in essence provide the sufficient field enhancement in their hot-spots for ultra-low concentration spectral analysis [11-13]. In practice, however, the reported values for the average field enhancement in these structures are fairly modest due to the large variation of the field profile over the sensing area. Moreover, in the case of large biomolecules, the non-uniformity of the enhancement profile could result in the unrepeatability of the experiments due to substantial size of the biomolecules relative to the hot-spots. The conventional route to SERS is to use the localized surface plasmon polaritons (SPP) in an array of isolated nanoantennas. The high absorption (and scattering) cross section of the localized SPPs near the resonance frequency results in a large enhancement of the Raman signal collected from the molecules located in the intense hot-spots. However, there is a practical limit to the intrinsic nanoantenna cross-sections, determined by the topology and material properties. A promising approach to improve the SERS enhancement is to use the sharp and asymmetric Fano-type resonances [14-22] that occur in coupled periodic plasmonic nanostructures due to the beating between the scattering caused by the periodicity of the nanostructure (i.e. Bragg effect) and the local nanoantenna scattering. These Fano-type resonant features have recently found some interesting applications. For instance, ultrasensitive refractive index sensing has been demonstrated using these resonant modes, taking advantage of their high sensitivity to any small perturbation in the environment [23,24]. More importantly, these collective resonant features are accompanied by plasmonic waves propagating across the structure, called lattice plasmons (LPs). Previously, LPs inside a nanoantenna array were used to enhance the stimulated emission [25,26] by increasing the local density of state (LDOS) at the band-edge, and hence the emission rate according to Fermi's golden rule [27,28]. However, this effect is very narrow-band and is not suitable for the enhancement of the wide-band Raman emission spectra.

Our approach, in this paper, is to utilize the band-edge LPs at the excitation (or pump) wavelength to increase the net absorption cross section of the array (to achieve more efficient pumping). The interaction between adjacent unit-cells of the bilayer array, at the LP band-edge, slightly extends the localized SPPs in space and forms a Bloch mode with near-zero group velocity that can be coupled more efficiently with the incoming light. The SPP modes of the nanostructure are adjusted to maximize the overall emission cross section, as well. Furthermore, the two interacting layers of the nanostructure create intense hot-spots in the vertically oriented gaps (over the dielectric nanopillars), which are coated selectively to form the sensing area. These two provisions collaboratively result in a large SERS enhancement over a large bandwidth, rivaling and potentially surpassing most nanofabricated SERS arrays. The rather uniform distribution of the SERS enhancement over the sensing area renders the structure particularly suitable for large biomolecules, such as proteins.

## Nanostructure Design and Fabrication

An array of gold nanodisks, stacked on an array of nanoapertures via supporting dielectric nanopillars, constitutes our nanostructure, illustrated in Fig. 1a and b. The dielectric nanopillars, composed of hydrogen-silsesquioxane (HSQ), act as the sensing area in SERS measurements (see the SI text for details of the fabrication method). In order to maximize the light-matter interaction, the periodicity of the structure is selected such that a band-edge LP with near-zero group velocity is induced at the close vicinity of the excitation wavelength, as a direct consequence of the lateral coupling between the nanoantennas. Additionally, the structure is designed to confine the localized SPP modes at the Fano resonance wavelength in the vertical gap between the two layers, providing a large and fairly uniform enhancement profile over the dielectric nanopillars (Fig. 2b and c). We found that the highest enhancement factor occurs, when the LP band-edge coincides with the resonance peak of a localized SPP resonance peak of the individual nanoantennas.

We performed an extensive search over the domain of the design parameters (i.e. the nanopillar radius and the lattice constant) through the full wave simulation of the structure using three-dimensional finite-difference time-domain (FDTD) method (Lumerical Inc.) to design the structure with this requirement, and the optimal design parameters are shown in Fig. 2a.

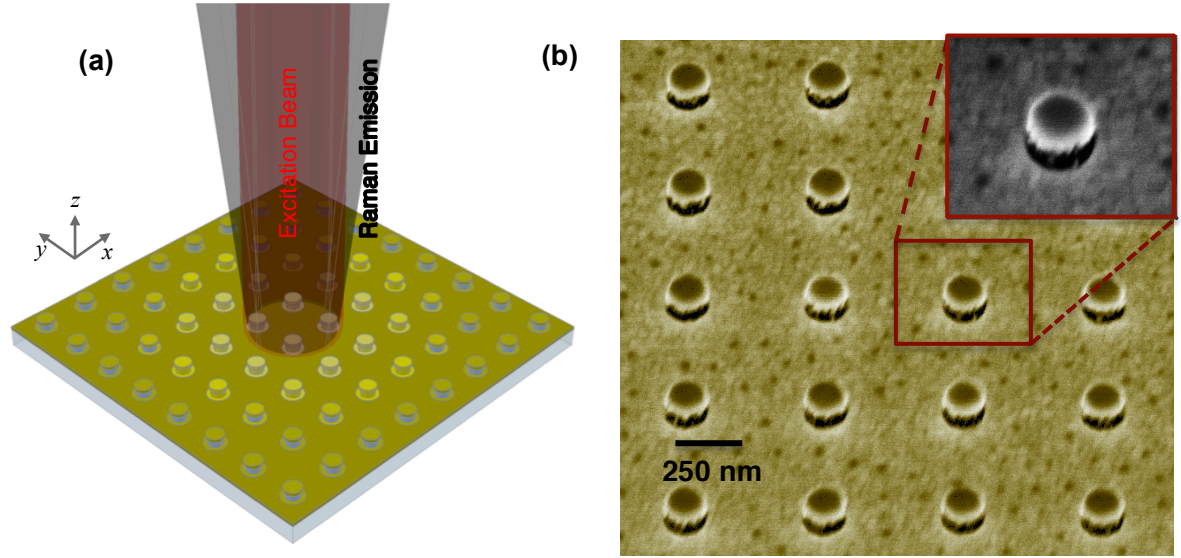

**Fig. 1.** *Band-edge bilayer plasmonic nanostructure. (a) Schematic view. (b) False-colored scanning electron microscope (SEM) image of a nanoantenna array (lattice constant of 540 nm and nanopillar radius of 80 nm in the layout) with 40° tilt angle revealing the mushroom-shaped topology of the nanoantennas with rims slightly extended following electron beam deposition.*

The nanostructure shown in Fig. 2a has three distinct localized SPP modes: a primarily disk mode at 668 nm (LSPR$_1$) and two vertical gap modes at 781 nm (LSPR$_2$) and 937 nm (LSPR$_3$), in which most of the energy is confined in the vertical gap between the nanoantenna and the nanoaperture (See Figure 2d-f). A thorough study of the plasmonic properties of these localized modes is provides in SI Text. Both of the two gap modes (LSPR$_2$ and LSPR$_3$) provide a large field enhancement with relatively uniform distribution over the dielectric surface. Within the range of SERS measurements (wavelengths between 797 nm and 931 nm), the second and third resonant modes both contribute to the overall Raman enhancement. In addition, LSPR$_2$, at 781 nm, coincides with the band-edge of the second LP mode (LP$_2$), near the pump wavelength (785 nm), Figure 3. This Fano-type plasmonic mode is very narrow band, but it has a large absorption cross section and field enhancement at the resonance peak. Hence, LSPR$_2$ is an excellent candidate for enhancing the narrow-band pump signal, at 785 nm wavelength. The second gap mode has a considerably larger bandwidth and contributes to the enhancement of the emitted Raman signal.

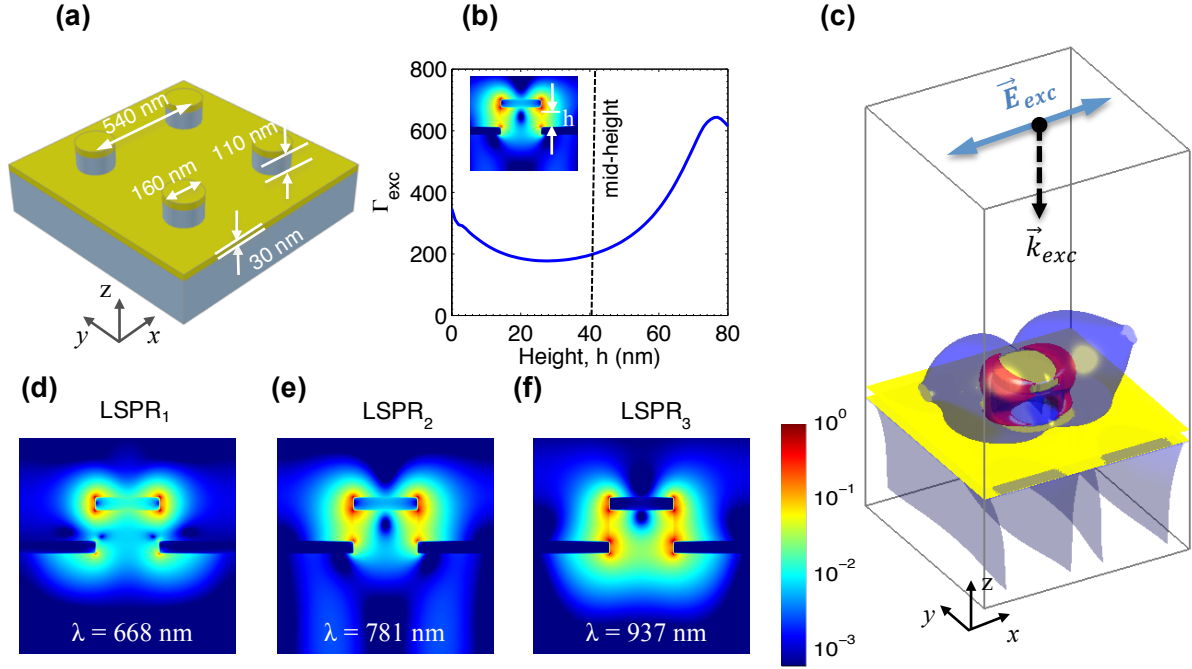

**Fig. 2.** *Distribution of the electrical field in a unit-cell of the nanostructure.* (a) *Geometrical parameters of the periodic array.* (b) *Distribution of excitation field enhancement,* $\Gamma_{exc}$, *along the dielectric nanopillar at 5 nm radial distance, from the edge of the nanoapertures to the lower edge of the nanodisk (30 nm of the dielectric is covered by the bottom gold layer).* (c) *Lateral and vertical coupling inside the nanostructure; the red surface (25% of the peak* $\Gamma_{exc}$*) shows the lateral coupling between the two layers of the nanostructure in an individual unit-cell, whereas the blue surface (5% of the peak* $\Gamma_{exc}$*) portrays the slightly extended Bloch mode (LP mode)*

*resulting in the lateral coupling between adjacent unit-cells. (d-f) Normalized electric-field distribution of the three localized SPPs (x-z cross-section) at 668, 781 and 937 nm, respectively.*

The structure shown in Fig. 2a has two LP modes, each one with a band-edge at normal incidence that can be adjusted in a wide frequency range by changing the lattice constant (nanopillar radius has a minor effect on the LP band-edge). On the other hand, the resonance peaks of the SPP modes of this structure can also be adjusted in a wide frequency range by changing the nanopillar radius and to some extent the lattice constant. Thus, the proposed structure can be adjusted to operate within a wide frequency gamut for the simultaneous enhancement of the excitation and emission signals (See the SI text). In our structure, the second LP band-edge (point A in Fig. 3) is used, as it can be excited more efficiently with the normally incident light and can be adjusted more easily.

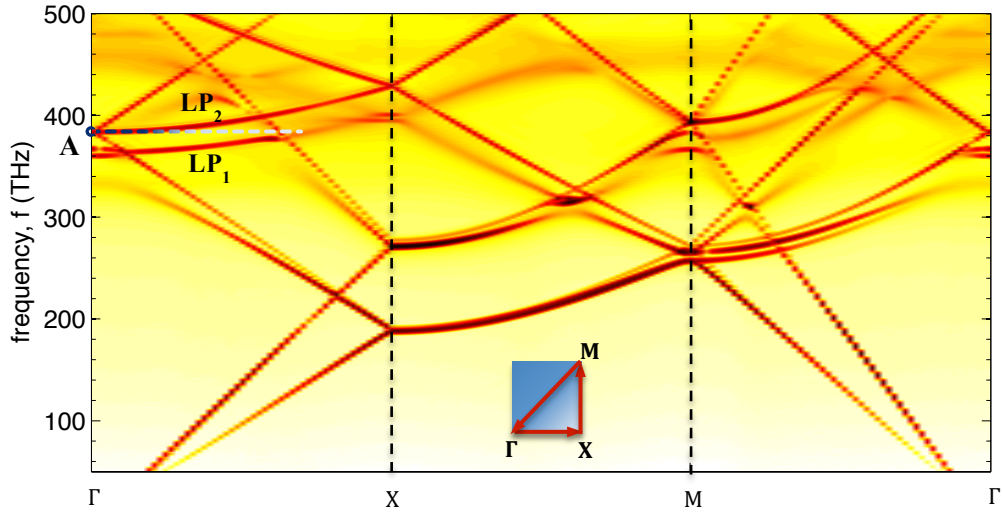

**Fig. 3.** *Full band-diagram of the nanostructure shown in Fig. 2a; LP$_1$ and LP$_2$ denote the LP modes, and point A shows the LP$_2$ band-edge, coincident with the second localized SPP (LSPR$_2$)*

*in Fig. 2e, the 2D Brillouin zone used in the calculation of the band-diagram is shown in the figure inset (this band-diagram calculation is also performed using 3D FDTD method).*

To assess the performance of our structure for sensing, we have calculated the local excitation and emission field enhancement spectra, $\Gamma_{exc}$ and $\Gamma_{em}$ respectively, using separate 3D FDTD simulations. Fig. 4a shows the $\Gamma_{exc}$ under the normal incidence (defined as the squared value of the local electric field under plane wave excitation with unit amplitude) at two fixed positions (points A and B in Figure 4a). The emission field enhancement ($\Gamma_{em}$), also known as the Purcell factor, for the three field components at point A in Fig. 4a is shown in Fig. 4b. In order to calculate the spontaneous emission, we have approximated the excited molecules by an electric dipole located at a fixed position point A (approximate position of the proteins attached to the nanopillars), and we have estimated the enhancement factor using the method proposed by Xu et al [29] (detailed calculation method described in the Appendix). In contrast to most reported works in plasmonic sensing, we use the dielectric surface of the nanopillars for the immobilization of the target molecules to lower the amount of analyte necessary for the coating of the nanostructure; an important consideration in many biosensing applications. The results shown in Fig. 4a and b depict close to the worst case for (i.e. the smallest) $\Gamma_{exc}$ and $\Gamma_{em}$ as the point A in Fig. 4a is located close to the minimal point of electric field distribution at 785 nm and 937 nm (Figure 2e and f). Nevertheless, the total enhancement factor in this worst-case scenario is comparable to the maximum enhancement factor in bowtie nanoantennas, which are one of the best performing nanofabricated nanoantennas, with air gaps as small as 50 nm [30-32]. The enhancement factor in our structure is higher at points that are closer to the surface of the nanopillar and / or are closer to the top nanodisk or the bottom nanoaperture. Most notably, the high Raman enhancement is present all over the sensing area, whereas in bowtie nanoantennas, the SERS hotspot is concentrated at a very small region between the two triangles of the nanostructure.

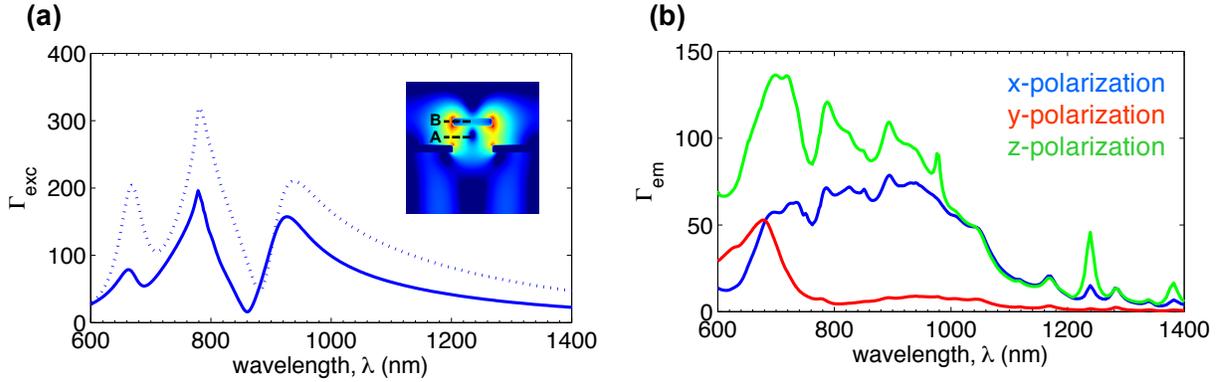

**Fig. 4.** *Excitation and emission enhancement spectra (simulation). (a) Excitation field enhancement ($\Gamma_{exc}$); the solid curve shows the enhancement profile at the mid-height of the nanopillar with 5 nm radial distance (point A in the inset), and the dotted curve shows the enhancement at the mid-height of the top nanodisk (point B) for the sake of comparison. (b) Enhancement factor of the spontaneous Raman emission ($\Gamma_{em}$) at point A in Fig. 4a for the x,y and z components of the electric field shown by blue, red and green curves, respectively.*

## Results and Discussion

To demonstrate the high sensitivity of our device, nanoantenna arrays with different radii and lattice constants were fabricated on a silicon wafer with a thick thermally grown $SiO_2$ layer on top. A medium-sized protein, streptavidin (53 kDa) was coated on the dielectric nanopillars by immersing the sample in an aquatic solution of the protein with the controlled concentration of 100 nM. Prior to the immobilization of streptavidin, the dielectric surface was functionalized using a process involving two self-assembled monolayers (SAMs), a layer of 3-aminopropyl-triethoxysilane (APTES) to provide free amine groups and and a second SAM of NHS-biotin, as the linker to trap the protein molecules.

The SERS spectrum of each array was collected using a near-infrared excitation laser at 785 nm. Fig. 5a shows the Raman spectra for arrays with the lattice constants varying from 500 nm to 580 nm and with the fixed pillar radius of 80 nm (in the layout). As expected from the simulations (see Fig. 3), the best Raman signal was acquired from the array with the lattice constant of 540 nm, Fig. 5a. The variation of the SERS signal with the nanopillar radius, at a fixed lattice constant is less prominent with the strongest signal acquired from the array with 80 nm radius nanopillars; Fig. 5b. Our simulations had predicted that the band-edge of $LP_2$ should coincide with the second localized SPP resonance in the array with 80 nm nanopillars and the periodicity of 540 nm, which was confirmed with the experiments.

It should be noted that fabricated nanoantennas have slightly different topology, from the ideal structure shown in Fig. 2.a. Close inspection by scanning electron microscopy (SEM), has revealed that the fabricated nanoantennas are mushroom-shaped (Fig. 1b) with rims slightly extended outside the supporting nanopillars due to the nature of the electron beam deposition. This should also affect the geometry of the nanoapertures at the bottom of the nanopillars. Despite this non-ideal shape, a good agreement between the theoretical prediction and the experiment was observed, suggesting that performance of our nanostructure is not sensitive to the impact of the fabrication imperfections on the topology of the nanoantennas.

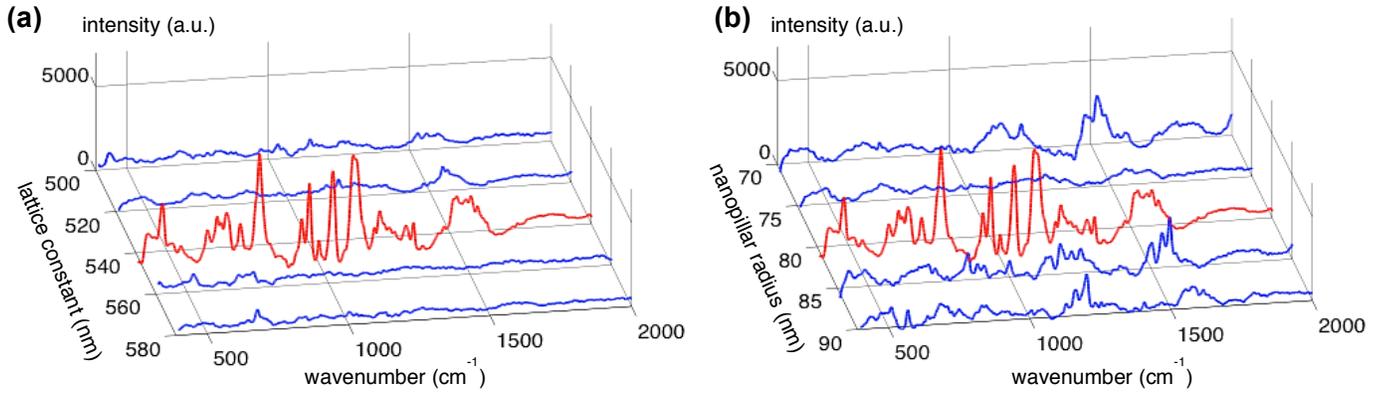

**Fig. 5.** *Comparison between the SERS spectra acquired from arrays with varying lattice constants and nanopillar radii. (a) SERS spectra acquired from the array with nanopillar radius of 80 nm and lattice constant varying between 500 nm to 580 nm. (b) SERS spectra acquired from the arrays with a fixed lattice constant of 540 nm and nanopillar radii varying between 70 nm and 90 nm; in both figures the red curve shows the SERS spectrum of the optimal array.*

## Conclusion

In summary, we presented here a novel bilayer plasmonic nanostructure for chip-scale SERS-based spectroscopic analysis of large biomolecules. Through the optimization of the horizontal coupling of the nanoantennas in the array, we were able to drastically improve the Raman scattering cross section of the nanoantenna array. The vertical coupling between the two layers further increases the energy confinement in the vertical gaps and provides a more uniform enhancement profile over the dielectric nanopillars. By opting for the dielectric nanopillar as the immobilization surface, the target molecules are more efficiently excited and all contribute significantly to the overall collected Raman signal. In other words, all the target molecules are bound to the surface coating at the hot-spots. These two effects collectively result in a large improvement in the overall efficiency of the SERS-based assay.

In experiments, we have successfully acquired the SERS spectrum of streptavidin, as an example of a large biomolecule, at the concentration of 100 nM using the optimized nanoantenna array, which to the best of our knowledge, shows a five-fold improvement compared to the previously reported plasmonic works [33]. Although we have used radially symmetric nanoantennas in this work to keep the sensitivity of the array to polarization as low as possible, the idea of using LP modes to increase the scattering cross section can be applied to other nanoantenna geometries to achieve even higher SERS enhancements. Our fabrication process is a great advantage in this regard, since higher aspect ratios and smaller gaps can be achieved using this method as compared to fabrication processes based on the lift-off or ion-beam milling.

## Appendix

In this short appendix, we describe the numerical methods for the calculation of the optical properties of the localized SPP modes, as well as the calculation of the excitation and emission enhancement factors. As seen from Fig. 4a, the LSPR$_2$ resonance lineshape is strongly altered due to the Fano resonance, and can no longer be modeled by a single Lorentzian. The resonance lifetimes, $\tau_{res}$ was estimated by fitting an exponential function to the envelope of the decaying electric field at a point near the nanoantenna (point A in Fig. 4a). Another important parameter of the localized SPP modes is the effective mode volume, $V_{eff}$, defined as Eq. 1. In the calculation of effective mode volumes, we used a modified definition for the energy density, $u_E$, (Eq. 2), proposed for lossy and dispersive material such as Au [34].

$$V_{eff} = \frac{\int u_E \, dv}{\max\left(u_E\right)} \ . \tag{1}$$

$$u_E = \frac{\varepsilon_0}{2}\left(\varepsilon_r + \frac{2\omega\varepsilon_i}{\gamma}\right)|E|^2 \ . \tag{2}$$

In Eq. 2, $\varepsilon_r$ and $\varepsilon_i$ are the real and imaginary part of the relative permittivity from the Drude model with the damping constant of $\gamma$, fitted to the Johnson and Christy's empirical data.

**Table 1.** *Optical properties of the localized SPP modes, calculated resonance lifetime, $\tau_{res}$ and effective mode volume, $V_{eff}$, for the three localized SPP modes (LSPR$_1$, LSPR$_2$ and LSPR$_3$).*

|  | $\lambda_{res}$ | $\tau_{res}$ | $V_{eff}$ |
|---|---|---|---|
| **LSPR₁** | 668 nm | 13.04 *fsec* | $2.04 \times 10^{-4} \ \lambda^3$ |
| **LSPR₂** | 781 nm | 16.50 *fsec* | $3.64 \times 10^{-4} \ \lambda^3$ |
| **LSPR₃** | 937 nm | 15.63 *fsec* | $8.04 \times 10^{-5} \ \lambda^3$ |

The emission rate enhancement factor, $\Gamma_{x,em}$, $\Gamma_{y,em}$ and $\Gamma_{z,em}$, were calculated by integrating the emitted power of the a dipole (placed at point A in Fig. 4a) within a surrounding closed surface in the presence of the nanostructure ($P_{i,struct}$), divided by the emitted power in vacuum ($P_{i,vac}$), as follows.

$$\Gamma_{i,em} = \frac{P_{i,struct}}{P_{i,vac}}, \quad i = x, y \text{ and } z.$$  (3)

## Materials and Methods

### Fabrication of the nanostructure

The nanoantenna arrays were fabricated on a $SiO_2$ wafer (4 μm of thermally grown $SiO_2$ on top of a silicon wafer) to minimize the fluorescence background. The oxide wafer was first covered by 110 nm of hydrogen-silsesquioxane (HSQ), which is a negative electron-beam resist and a spin-on dielectric material. Then, the thin HSQ layer was patterned using electron-beam lithography to form the nanopillars, and 100 by 100 arrays with different nanopillar radii and lattice constants were formed on the oxide surface. Finally, 3 nm of titanium and 30 nm of gold were deposited using electron-beam deposition to form the self-aligned bilayer nanostructure using a single lithography step.

### Sample preparation

Attachment sites for streptavidin were prepared by forming two stacked SAMs on the dielectric nanopillars. First, the sample was soaked in a 4% solution of APTES (Sigma Aldrich) in pure ethanol at room temperature for 2 hours to form frees amine groups. After washing the sample in ethanol, it was soaked again in a 1 mg/ml solution of NHS-biotin (Pierce Biotechnology) in dimethyle solfoxide (DMSO) for an additional 2 hours, also at room temperature, forming a second monolayer with biotin attachment sites for trapping streptavidin. Finally, the sample was soaked in an aquatic solution of streptavidin with the controlled concentration of 100 nM at 4 °C for 1 hour.

**Acquisition of the SERS spectra**

The Raman spectra were acquired using a commercial Raman spectrometer (Renishaw Inc.) coupled to an inverted microscope. A 50X objective lens with the numerical aperture of 0.75 was used to focus about 1 mW power from a near infrared laser (wavelength: 785 nm) on the sample. The Raman signal was acquired in the range of 400 to 2000 cm$^{-1}$ with 3 min acquisition time.

# Acknowledgment

This work was supported Defense Advanced Research Projects Agency (DARPA) under Contract No. HR 0011-10-1-0075. The authors would like to thank F. Ghasemi for sharing his surface functionalization recipe. They also thank S. R. Panikkanvalappil, M. Mahmoud and M. El-Sayed for providing access to their setup and helping with the Raman measurements.

# References

[1]     Kneipp K, et al (1997) Single molecule detection using surface-enhanced Raman scattering (SERS). *Phys Rev Lett* 78(9): 1667.

[2]     Nie S, Emory S R (1997) Probing single molecules and single nanoparticles by surface-enhanced Raman scattering. *Science* 275(5303): 1102-1106.

[3]     Barnes W L, Dereux A, Ebbesen T W (2003) Surface plasmon subwavelength optics. *Nature* 424(6950): 824-830.

[4]     Lal S, Link S, Halas N J (2007) Nano-optics from sensing to waveguiding. *Nat Photonics* 1(11): 641-648.

[5]     Novotny L, Van Hulst V (2011) Antennas for light. *Nat Photonics* 5(2): 83-90.

[6]     Schuller J A, et al (2010) Plasmonics for extreme light concentration and manipulation. *Nat Mater* 9(3): 193-204.

[7]     Tuma R (2005) Raman spectroscopy of proteins: from peptides to large assemblies. Journal of Raman Spectroscopy. *J Raman Spectrosc* 36(4): 307-319.

[8]     Callender R, Deng H (1994) Nonresonance Raman difference spectroscopy: a general probe of protein structure, ligand binding, enzymatic catalysis, and the structures of other biomacromolecules. *Annu Rev Biophys Biomol Struct* 23(1): 215-245.

[9]     Thomas G J, Jr (1999) Raman spectroscopy of protein and nucleic acid assemblies. *Annu Rev Biophys Biomol Struct* 28(1): 1-27.

[10]    Gniadecka, M, et al (2004) Melanoma diagnosis by Raman spectroscopy and neural networks: structure alterations in proteins and lipids in intact cancer tissue. *J Invest Dermatol* 122(2): 443-449.

[11]    Talley C E, et al (2005) Surface-enhanced Raman scattering from individual Au nanoparticles and nanoparticle dimer substrates. *Nano Lett* 5(8): 1569-1574.

[12]    Jackson J B, Westcott S L, Hirsch L R, West J L, Halas N J (2003) Controlling the surface enhanced Raman effect via the nanoshell geometry. *Appl Phys Lett* 82(2): 257-259.

[13]    Li J F, et al (2010) Shell-isolated nanoparticle-enhanced Raman spectroscopy. *Nature* 464(7287): 392-395.

[14]    Auguié B, Barnes W L (2008) Collective resonances in gold nanoparticle arrays. *Phys Rev Lett* 101(14): 143902.

[15]    Luk'yanchuk B, et al (2010) The Fano resonance in plasmonic nanostructures and metamaterials. *Nat Mater* 9(9): 707-715.

[16]    Hicks E M, et al (2005) Controlling plasmon line shapes through diffractive coupling in linear arrays of cylindrical nanoparticles fabricated by electron beam lithography. *Nano Lett* 5(6): 1065-1070.

[17]    Stuart H R, Hall D G (1998) Enhanced dipole-dipole interaction between elementary radiators near a surface. *Phys Rev Lett* 80(25): 5663.

[18]    Lamprecht B, et al (2000) Metal nanoparticle gratings: influence of dipolar particle interaction on the plasmon resonance. *Phys Rev Lett* 84(20): 4721-4724.

[19]    Giannini V, Vecchi G, Gómez Rivas J (2010) Lighting up multipolar surface plasmon polaritons by collective resonances in arrays of nanoantennas. *Phys Rev Lett* 105(26): 266801.

[20]    Zhou W, Odom T W (2011) Tunable subradiant lattice plasmons by out-of-plane dipolar interactions. *Nat Nanotechnol* 6(7): 423-427.

[21]    Väkeväinen A, et al (2013) Plasmonic surface lattice resonances at the strong coupling regime. *Nano Lett* 14(4): 1721-1727.

[22]    Chu Y, Schonbrun E, Yang T, Crozier K B (2008) Experimental observation of narrow surface plasmon resonances in gold nanoparticle arrays. *Appl Phys Lett* 93(18): 181108.

[23]    Adato R, et al (2009). Ultra-sensitive vibrational spectroscopy of protein monolayers with plasmonic nanoantenna arrays. *Proc Natl Acad. Sci USA*, 106(46), 19227-19232.

[24]    Malyarchuk V, Stewart M E, Nuzzo R G, Rogers J A (2007) Spatially resolved biosensing with a molded plasmonic crystal. *Appl Phys Lett* 90(20): 203113.

[25]    Zhou W, et al (2013) Lasing action in strongly coupled plasmonic nanocavity arrays. *Nat Nanotechnol* 8(7): 506-511.

[26]    Van Beijnum F, et al (2013) Surface plasmon lasing observed in metal hole arrays. *Phys Rev Lett* 110(20): 206802.

[27]    Novotny L, Hecht B (2012) *Principles of nano-optics*. Cambridge university press.

[28]    Baba T (2008) Slow light in photonic crystals. *Nat. Photonics* 2 (8): 465-473.

[29]    Xu Y, Lee R K, Yariv A (2000) Quantum analysis and the classical analysis of spontaneous emission in a microcavity. *Phys Rev A* 61(3): 033807.

[30]    Schuck P J, Fromm D P, Sundaramurthy A, Kino G S, Moerner W E (2005) Improving the mismatch between light and nanoscale objects with gold bowtie nanoantennas. *Phys Rev Lett.* 94(1): 017402.

[31]    Kinkhabwala A, et al (2009) Large single-molecule fluorescence enhancements produced by a bowtie nanoantenna. *Nat Photonics* 3(11): 654-657.

[32]    Fromm D P, Sundaramurthy A, Schuck, P J, Kino G, Moerner W E (2004) Gap-dependent optical coupling of single "bowtie" nanoantennas resonant in the visible. *Nano Lett* 4(5): 957-961.

[33]    Galarreta B C, Norton P R, Lagugné-Labarthet F (2011) SERS Detection of Streptavidin/Biotin Monolayer Assemblies. *Langmuir* 27(4): 1494-1498.

[34]    Sauvan C, Hugonin J P, Maksymov I, Lalanne P (2013) Theory of the spontaneous optical emission of nanosize photonic and plasmon resonators. *Phys Rev Lett* 110(23): 110237401.

**Figure Legends**

**Fig. 1.** Band-edge bilayer plasmonic nanostructure**.** (a) Schematic view. (b) False-colored scanning electron microscope (SEM) image of a nanoantenna array (lattice constant of 540 nm and nanopillar radius of 80 nm in the layout) with 40° tilt angle revealing the mushroom-shaped topology of the nanoantennas with rims slightly extended following electron beam deposition.

**Fig. 2.** Distribution of the electrical field in a unit-cell of the nanostructure**.** (a) Geometrical parameters of the periodic array. (b) Distribution of excitation field enhancement, $\Gamma_{exc}$ , along the dielectric nanopillar at 5 nm radial distance, from the edge of the nanoapertures to the lower edge of the nanodisk (30 nm of the dielectric is covered by the bottom gold layer). (c) Lateral and vertical coupling inside the nanostructure; the red surface (25% of the peak $\Gamma_{exc}$ ) shows the lateral coupling between the two layers of the nanostructure in an individual unit-cell, whereas the blue surface (5% of the peak $\Gamma_{exc}$ ) portrays the slightly extended Bloch mode (LP mode) resulting in the lateral coupling between adjacent unit-cells. (d-f) Normalized electric-field distribution of the three localized SPPs (x-z cross-section) at 668, 781 and 937 nm, respectively.

**Fig. 3.** Full band-diagram of the nanostructure shown in Fig. 2a; $LP_1$ and $LP_2$ denote the LP modes, and point A shows the $LP_2$ band-edge, coincident with the second localized SPP ($LSPR_2$) in Fig. 2e, the 2D Brillouin zone used in the calculation of the band-diagram is shown in the figure inset (this band-diagram calculation is also performed using 3D FDTD method).

**Fig. 4.** Excitation and emission enhancement spectra (simulation). (a) Excitation field enhancement ($\Gamma_{exc}$); the solid curve shows the enhancement profile at the mid-height of the nanopillar with 5 nm radial distance (point A in the inset), and the dotted curve shows the enhancement at the mid-height of the top nanodisk (point B) for the sake of comparison. (b) Enhancement factor of the spontaneous Raman emission ($\Gamma_{em}$) at point A in Fig. 4a for the x,y and z components of the electric field shown by blue, red and green curves, respectively.

**Fig. 5.** Comparison between the SERS spectra acquired from arrays with varying lattice constants and nanopillar radii. (a) SERS spectra acquired from the array with nanopillar radius of 80 nm and lattice constant varying between 500 nm to 580 nm. (b) SERS spectra acquired from the arrays with a fixed lattice constant of 540 nm and nanopillar radii varying between 70 nm and 90 nm; in both figures the red curve shows the SERS spectrum of the optimal array.

**Supporting Information**

# 1.    Fabrication process

All nanoantenna arrays were fabricated on a common silicon substrate with a 4 μm-thick layer of thermal oxide on top. The overall size of the 100 by 100 arrays varies between 50 to 58 μm, depending on the lattice constant. The thermal oxide was grown on a silicon wafer by wet oxidation in an oxidation furnace at 1100 °C. Consequently, the $SiO_2$ substrate was covered by a 110 nm-thick uniform layer of hydrogen-silsesquioxane (HSQ) by spin-coating XR-1541 resist (Dow Corning), containing a 6% solution of HSQ in methyl isobutyl ketone (MIBK), with the spinning speed of 6000 rpm for 60 seconds and 2 seconds of ramping time. HSQ is a spin-on dielectric material, used in semiconductor industry as an inter-level dielectric in multilevel structures, and it is also a negative-tone electron beam resist that can be patterned by electron beam lithography (EBL) [1]. Upon exposure with electron-beam, HSQ converts to a chemically stable amorphous dielectric film. To pattern the HSQ layer, the sample was first prebaked at 90 °C for 3 min. Then, EBL was performed, using JEOL JBX-9300FS EBL, with a dosage of 3720 μC/cm$^2$. In order to resolve the patterns and form the nanopillars, the sample was immersed in a 25% aquatic solution of tetramethylammonium hydroxide (TMAH) for 30 sec, and was rinsed by deionized water for 5 minutes. The two layers of the nanostructure were finally formed by depositing 3 nm of titanium (Ti) as the adhesion layer and 30 nm of gold (Au) using electron-beam evaporation.

We measured the diameter of the nanopillars before and after electron beam deposition. As shown in Fig. S1a, the diameter of the HSQ nanopillars after EBL is smaller compared to their size in the prepared layout (132 nm versus 140 nm). Top SEM view of the nanodisks, following electron beam evaporation (Fig S1b), revealed that the diameter was increased relative to the dielectric nanopillars (138 nm vs. 132 nm). Moreover, the nanodisks are mushroom-shaped, with rims slightly extended outside the dielectric nanopillars (Fig. S1.c). The altered size and topology of the fabricated nanoantennas, relative to the prepared layout, necessitates an experimental sweep over the nanopillar diameter to find the array with optimal response.

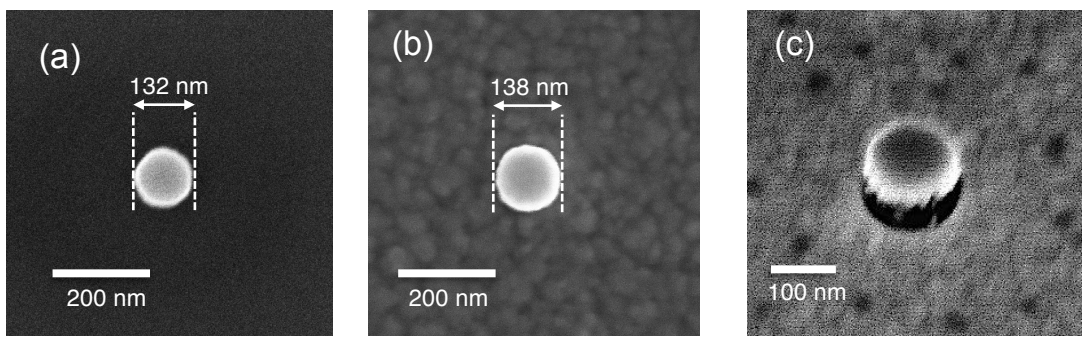

**Fig. S1.** *SEM images of one unit-cell before and after electron beam deposition. (a) Top SEM view of one HSQ nanopillar with the diameter of 140 nm in the layout. (b) Top SEM view of the mushroom-shaped nanodisks after E-beam evaporation. (c) Oblique SEM view of a nanodisk-nanoaperture pair with a 40 ° tilt angle.*

Next, we prepared the sample for the attachment of protein molecules. First, the sample was soaked in a 4% solution of 3-aminopropyltriethoxysilane (APTES), Sigma Aldrich, in pure ethanol for 4 hours at room temperature to silanize the dielectric nanopillars, making a self-assembled monolayer (SAM) with free amine groups. The sample was removed from the solution, rinsed with ethanol and baked on a hot plate at the temperature of 100 °C for 10 minutes, to stabilized the APTES monolayer. Then, a second SAM was formed, by immersing the sample in a solution of N-hydroxysuccinimido biotin (NHS-biotin), Thermo Scientific, in dimethyl sulfoxide (DMSO) with the concentration of 1 mg/ml, for another 4 hours at room temperature. NHS-biotin is a biotinyltation reagent and acts as the linker between the free amine groups and the proteins. Finally, we immersed the sample in an aquatic solution of the protein (streptavidin), Thermo Scientific, with the concentration of 100 nM for 1 hour at 4 °C. By lowering the temperature, we speed up the binding kinetics of the proteins to the Biotin monolayer.

## 2.     Band-diagram calculation

For the calculation of the bind-diagram, we have used the 3D finite difference time-domain (FDTD) method using a commercial software package (Numerical Inc.). The empirical data recorded by Johnson and Christy [2] was used to model the permittivity of the Au layers. The simulation area is laid down by setting perfectly matched layer (PML) boundary conditions in z-direction and Bloch boundary conditions in the x and y (coordinates shown if Fig. S.2a). The mesh size was kept at the fixed value of 4 nm in a 540 nm × 540 nm × 540 nm cubic area that was centered at the surface of the substrate (for the array with the lattice constant of 540 nm). Outside this high-resolution zone, a course adaptive mesh was used. A set of electric dipoles with random polarizations was placed inside the simulation area in order to excite all plasmonic and photonic modes. A set of time-domain probes with random locations was used to record the resonance response of the nanostructure for every value of the k-vector. Using, the harmonic inversion method, the output of all these time-domain probes are combined and the spectrum is obtained. To obtain the complete band-diagram, the k-vector was swept in the Brillouin zone, as shown in Fig. 3 in the paper.

Figure S2b shows the partial band-diagram of the array with the nanopillar radius of 80 nm and lattice constant of 540 nm, for $k_x$ between 0 and 0.5, $k_y = 0$, in the frequency range of 300 to 500 THz. The two straight lines, $L_1$ and $L_2$, are the light lines corresponding to air (folded at 381 THz) and $SiO_2$. Four confined modes can be identified in this figure. $LP_1$ and $LP_2$ are the two lattice plasmon modes of the bilayer periodic structure. The structure is tuned such that $LP_2$ coincides with the localized SPP of the bilayer nanoantenna at 381 THz. This lattice plasmon mode acts as a grating coupler and effectively increases the scattering cross-section of the nanoantennas. It should also be noted that although $LP_1$ is a propagative mode, at the excitation frequency, it has close-to-zero group velocity ($v_g$) and hence its energy remains contained in a small region near the excitation area. $SP_1$ and $SP_2$ are the two first-order surface plasmon modes of the bottom holey gold layer, corresponding to its interface with the $SiO_2$ substrate and air, respectively. These two SP modes are tangential to the air and $SiO_2$ light lines, as expected.

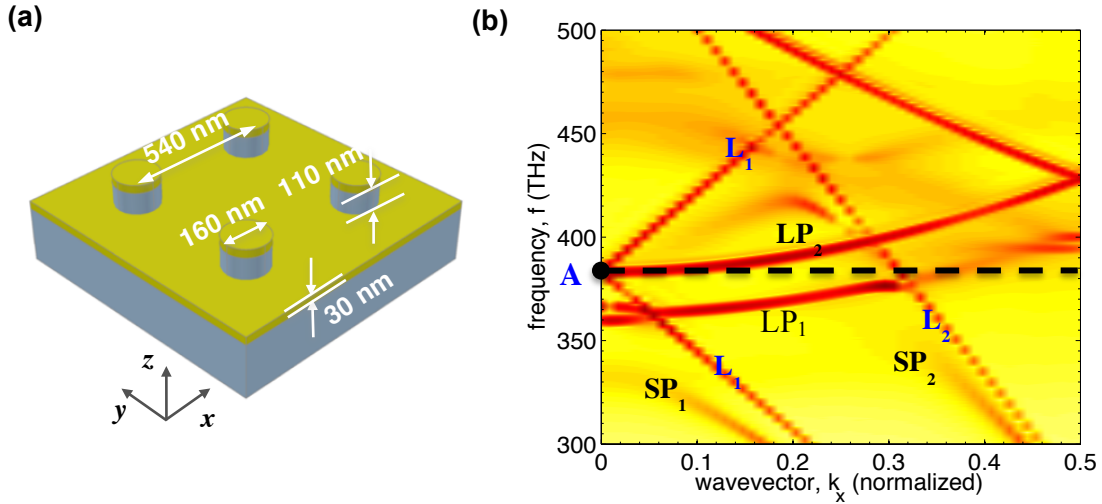

**Fig. S2.** *Band-diagram analysis of the nanostructure, (a) Geometrical parameters of the periodic array, (b) partial band-diagram for the array with lattice constant of 540 nm and nanopillar radius of 80nm, wave-vectors swept between points Γ and M in the Brillouin zone.*

Spectral Tuning of the LP$_2$ can be accomplished by adjusting the lattice constant, and to a lesser extent, the nanopillar radius. Figures S3a, b and c show the partial band-diagrams for three arrays with the lattice constants (a) of 510 nm, 540 nm and 570 nm, respectively and for a fixed nanopillar radius of 80 nm. As it can be seen from these figures, the LP$_2$ band-edge can readily be tuned in a wide range, by just changing the lattice constant.

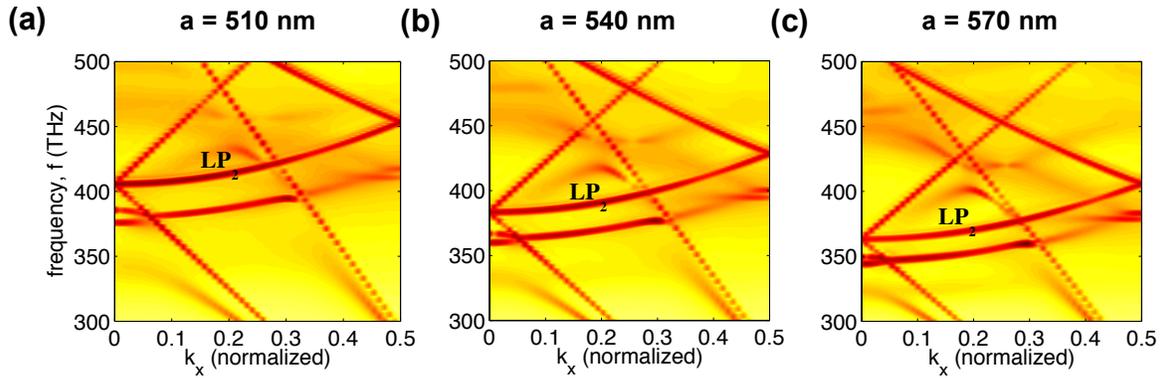

**Fig. S3.** *Spectral tuning of the LP$_2$ band-edge, (a), (b) and (c) partial band-diagrams for the arrays with the nanopillar radius of 80 nm and lattice constants of 510 nm, 540 nm and 570 nm, respectively.*

# 3.  Calculation of local electric field distribution profile, enhancement factors, and emission rates

The near-field enhancement spectra and electric field distribution profiles were calculated using separate FDTD simulations. A plane-wave excitation from the top was used to model the pump signal. The excitation enhancement spectrum, $\Gamma_{exc}$, was obtained by performing Fourier transform over the time-domain signal recorded by point monitors at the locations of interest.

The electric field profiles were also obtained by placing frequency-domain monitors at the resonance wavelengths.

To calculate the resonance lifetimes, $\tau_{res}$, we have excited the structure by a single frequency plane-wave source and switched off the source abruptly. By fitting an exponential function to the envelope of the decaying electric field at a point near the nanoantenna, we estimated the photon lifetimes. The mode volume, $V_{eff}$, was calculated using the electric field distribution profiles obtaind at three resonance wavelengths, and Eq. 1 in appendix [3].

Three other simulations were performed to calculate the spontaneous emission enhancement (Purcell factor), $\Gamma_{em}$, at point A for an electric dipole (modeling an excited molecule) with three different polarizations. Using the method descried by Xu et al [4,5], a dipole was placed at point A in the vicinity of a nanopillar inside a simulation area containing several periods of the structure and was assumed to be infinitely long in x and y directions. The emission rate enhancement factor, $\Gamma_{x,em}$, $\Gamma_{y,em}$ and $\Gamma_{z,em}$, were calculated by integrating the emitted power of the dipole within a surrounding closed surface in the presence of the nanostructure ($P_{i,struct}$), divided by the emitted power in vacuum ($P_{i,vac}$), as described in the appendix.

## 4.    SERS measurement and background rejection

The SERS spectra were acquired using a commercial Raman spectrometer (Renishaw Invia) coupled to a 1800 lines/mm grating, a CCD camera, and a 785 nm near-infrared laser with 100 mW maximum output power, coupled to a Leica microscope. The excitation laser was focused on the sample using a 50X objective lens (NA = 0.75) and the backscattered Raman emission was collected with the same objective. In SERS measurements, an optical attenuator with the optical density (OD) of 2 has been used in the excitation path to lower the excitation power to less than 1 mW in order to prevent thermal effects. The spectral resolution of the spectrometer is 0.94 cm⁻¹. The Raman spectra were acquired in the range of 400 to 2000 cm⁻¹ with an acquisition time of 3 minutes. The background fluorescence was removed from the collected signal by fitting the acquired data with a 9th order polynomial, as shown in Fig. S5a.

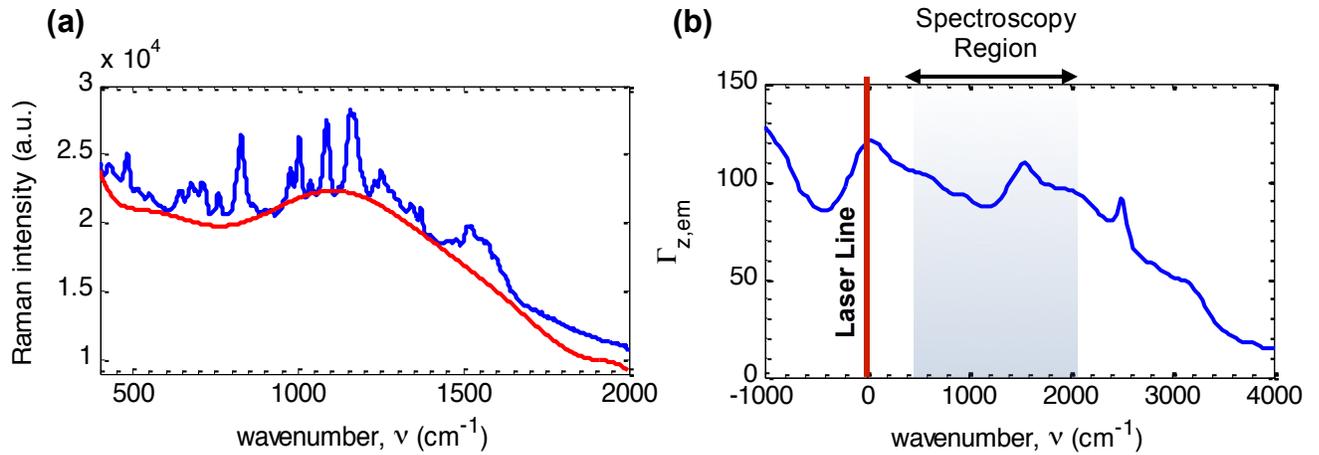

**Fig. S4.** *Background fluorescence rejection, (a) blue curve shows the SERS spectrum acquired from the optimal array (a = 540 nm, d = 160 nm), and the red curve is the 9th polynomial fitted to the SERS spectrum to remove the background fluorescence, (b) the calculated emission enhancement spectrum for a dipole polarized at z-direction; it is notable that the overall shape of the background fluorescence in the spectroscopy region roughly follows the calculated emission spectrum.*

# SI References

[1]     Coquand R, et al (2013) On the optimization of ebeam lithography using Hydrogen SilsesQuioxane (HSQ) for innovative self-aligned CMOS process. *ECS Trans* 53(3): 177-184.

[2]     Johnson P B, Christy R W (1972) Optical constants of the noble metals. *Phys Rev B* 6(12): 4370.

[3]     Maier S A (2006) Plasmonic field enhancement and SERS in the effective mode volume picture. *Opt Express* 14(5): 141957-1964.

[4]     Young Chul J, Briggs R M, Atwater H A, Brongersma M L (2009) Broadband enhancement of light emission in silicon slot waveguides." *Opt express* 17(9): 7479-7490.

[5]     Xu Y, Lee R K, Yariv A (2000) Quantum analysis and the classical analysis of spontaneous emission in a microcavity. *Phys Rev A*  61(3): 033807.

## SI Figure Legends

**Fig. S1.**    SEM images of one unit-cell before and after electron beam deposition. (a) Top SEM view of one HSQ nanopillar with the diameter of 140 nm in the layout. (b) Top SEM view of the mushroom-shaped nanodisks after E-beam evaporation. (c) Oblique SEM view of a nanodisk-nanoaperture pair with a 40 ° tilt angle.

**Fig. S2.**    Band-diagram analysis of the nanostructure, (a) Geometrical parameters of the periodic array, (b) partial band-diagram for the array with lattice constant of 540 nm and nanopillar radius of 80nm, wave-vectors swept between points $\Gamma$ and M in the Brillouin zone.

**Fig. S3.**    Spectral tuning of the $LP_2$ band-edge, (a), (b) and (c) partial band-diagrams for the arrays with the nanopillar radius of 80 nm and lattice constants of 510 nm, 540 nm and 570 nm, respectively.

**Fig. S4.**    Background fluorescence rejection, (a) blue curve shows the SERS spectrum acquired from the optimal array (a = 540 nm, d = 160 nm), and the red curve is the $9^{th}$ polynomial fitted to the SERS spectrum to remove the background fluorescence, (b) the calculated emission enhancement spectrum for a dipole polarized at z-direction; it is notable that the overall shape of the background fluorescence in the spectroscopy region roughly follows the calculated emission spectrum.